\documentclass[12pt]{article}
\usepackage{epsfig}

\def\etal{{\em et al.}}

\def\prl#1#2#3{{\it Phys. Rev. Lett.} {\bf #2} (#1) #3}
\def\prd#1#2#3{{\it Phys. Rev. D} {\bf #2} (#1) #3}
\def\plb#1#2#3{{\it Phys. Lett. B} {\bf #2} (#1) #3}
\def\vub{$| V_{ub}| $}
\def\vcb{$| V_{cb}| $}
\def\dec{\rightarrow}
\input psfig
\begin{document}
\begin{titlepage}
\large
\centerline{THE QUEST FOR THE CABIBBO KOBAYASHI MASKAWA MATRIX}
\vskip 2cm
\normalsize
\centerline{M. Artuso~\footnote{artuso@phy.syr.edu}}
\centerline{\it Syracuse University, Syracuse, NY 13244}
\vskip 4.0cm
\centerline{\bf ABSTRACT}
\vskip 1cm
A piece of the Standard Model presently undergoing intense experimental 
scrutiny  is
the Cabibbo Kobayashi Maskawa
matrix. Several different measurements are planned to enrich
the spectrum of experimental constraints and thus provide 
one of the most stringent tests of Standard Model
validity. The success of this program is closely related
to  theoretical progress in evaluating QCD matrix elements in
a non-perturbative regime, as
we need to extract fundamental quark properties from observations on 
decays involving hadrons.
This interplay between experimental and theoretical progress will be 
illustrated in the context of the present knowledge of the magnitudes 
of the quark mixing parameters $| V_{cb}| $ and $| V_{ub}|$.
\vfill
\end{titlepage}
\newpage
\section{Introduction}
In the framework of the Standard Model the gauge bosons, $W^{\pm}$, 
$\gamma$ and 
$Z^o$ couple to  
mixtures of the physical $d,~ s$ and $b$ states. This mixing is described
by the Cabibbo-Kobayashi-Maskawa (CKM) matrix:
\begin{equation}
V_{CKM} =\left(\begin{array}{ccc} 
V_{ud} &  V_{us} & V_{ub} \\
V_{cd} &  V_{cs} & V_{cb} \\
V_{td} &  V_{ts} & V_{tb}  \end{array}\right).
\end{equation}
Since the CKM matrix must be unitary, it can be expressed as a function of 
only four parameters. 
A commonly used approximate parameterization was originally proposed by 
Wolfenstein. 
It reflects the hierarchy between the magnitude of matrix elements 
belonging to different 
diagonals. In the form accurate to $\lambda ^3$ for the real part and 
$\lambda ^5$ for the imaginary part, it is given by:
\begin{equation}
\left({\begin{array}{ccc}
1-\lambda^2/2 &  \lambda & A\lambda^3(\rho-i\eta(1-\lambda^2/2)) \\
-\lambda &  1-\lambda^2/2-i\eta A^2\lambda^4 & A\lambda^2(1+i\eta\lambda^2) \\
A\lambda^3(1-\rho-i\eta) &  -A\lambda^2& 1  
\end{array}}\right).
\end{equation}

This matrix has already several experimental constraints \cite{pdg}, and
both the number of measurements and their accuracy
will see remarkable improvements in the
next few years and may eventually lead to evidence for
new physics. In order to accomplish this goal,
precision measurements are needed, as well as sophisticated theoretical
calculations able to evaluate QCD
matrix elements in a regime where non-perturbative effects are important. 
The interplay between theoretical and experimental errors will be
discussed with reference to the two quark mixing parameters $| V_{cb}|$
and $| V_{ub}| $.

\section{Experimental determination of the quark mixing parameter $|V_{cb}|$}
This parameter enables us to determine the Wolfenstein parameter $A$. 
It strongly affects the effectiveness of the $CP$ violation 
parameter in the $K^o-\bar{K}^o$ system, 
$\epsilon _K$~\cite{cpk}, 
in constraining $\rho$ and $\eta$, as 
 $\epsilon _K$ depends on $A^4$. 
$| V_{cb} |$ is 
measured by studying semileptonic decays of the type $B\dec X_c \ell
\nu$, where $X_c$ is a charmed hadron. Two approaches have 
been taken: 
an ``inclusive'' method, focusing on the lepton momentum
spectrum, and an ``exclusive'' method, studying a specific
channel, most notably the dominant decay $B\dec D^{\star} \ell \nu$. 

The decay $B\dec D^{\star} \ell \nu$ has received considerable attention, 
especially after the
theoretical development known as Heavy Quark Effective Theory (HQET), that 
has offered the opportunity to replace quark models of various nature with
an effective theory, exact in the limit of infinite quark masses and where
non-perturbative effects can be expressed in powers of $1/m_Q$.  In 
this approach, the
distribution $d\Gamma/dw$ is given by:
\begin{equation}
\frac{d\Gamma}{dw} = \frac{G_F^2| V_{cb}| ^2}{48\pi ^3}{\cal K}(w){\cal F}(w)^2,
\end{equation}
where $w$ is the inner product of the $B$ and $D^{\star}$
meson 4-velocities, ${\cal K}(w)$ is a known phase space 
factor and the form factor 
${\cal F}(w)$ is generally expressed as the product of a normalization
factor ${\cal F}(1)$ and a shape function, $g(w)$,
constrained by dispersion relations \cite{grinstein}.  There are 
several different corrections to the infinite
mass value ${\cal F}(1)=1$:
\begin{equation}
{\cal F}(1) =\eta _{QED}\eta _A \left[ 1 + \delta _{1/m^2} + ...\right]
\end{equation}
Note that, by virtue of Luke's theorem \cite{luke},
the first term in the non-perturbative expansion in powers of $1/m_Q$
vanishes. $QED$ corrections up to leading logarithmic order give $\eta
_{QED}\approx 1.007$, $QCD$ radiative corrections to two loops give $\eta _A =
0.960\pm 0.007$ and different estimates of the $1/m^2$ corrections,
involving
terms proportional to $1/m_b^2$, $1/m_c^2$ and $1/m_bm_c$, give an
average value $\delta _{1/m^2} = -0.55 \pm 0.035$ \cite{babarph}.
These corrections give ${\cal F}(1)=0.913 \pm 0.007 \pm 0.024 \pm 0.011$,
 where the first error represents uncertainties in radiative
corrections, the second uncertainties in $1/m^2$ corrections and the
last one is related to higher order power corrections. Adding
the errors linearly, we get ${\cal F}(1)=0.913 \pm 0.042$, that will be used to
extract $|V_{cb}|$ from data.  
The preliminary value from a quenched Lattice HQET calculation 
is $0.931\pm 0.035$ \cite{kronfeld}, in
good agreement with the previous estimate. This is a situation that is 
good: several different approaches have been used to evaluate a crucial input
parameter and they have close central values and comparable 
uncertainties.  

\begin{figure}[htb]
\centerline{\epsfig{figure=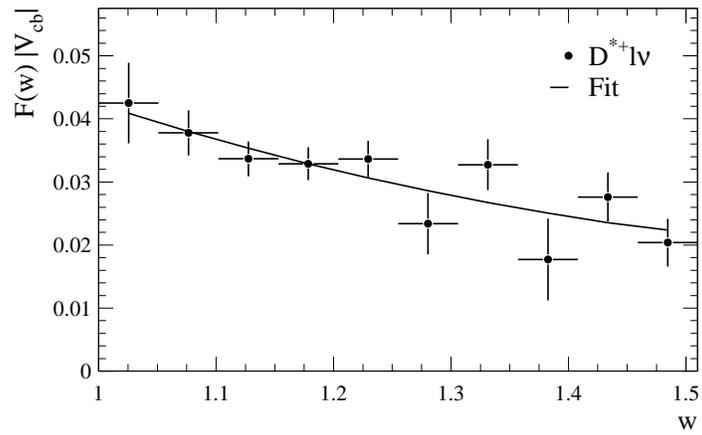,width=3.8in}}
\caption{\label{dslnu-cleo} $\overline{B}^o\to D^{*+}\ell^-\bar{\nu}$ from CLEO. 
The data have been
fit to a functional form suggested by dispersion relations \cite{grinstein}. 
The abscissa gives the
value of the product $|F(w)\cdot V_{cb}|$.}
\end{figure}

Experiments determine the product ${\cal F}(1) |V_{cb}|$ by
fitting the measured ${d\Gamma}/{dw}$ distribution. Fig.~\ref{dslnu-cleo}
shows the recent CLEO measurement \cite{vcb-cleo} of ${\cal F}(w)| V_{cb} |$ as a
function of $w$, based on a data sample of 3.33 millions $B\bar{B}$
pairs. They obtain ${\cal F}(1)|V_{cb}| = 0.0424\pm
0.0018\pm 0.0019$, where the first error 
is statistical and the second is
systematic, dominated by the uncertainty in slow $\pi$ finding efficiency.
The LEP experiments have also studied ${\cal F}(w)| V_{cb} |$ with
different experimental approaches \cite{lep-vcb}. Table~\ref{vcb-excl}
summarizes the data available so far. The agreement between these measurements
is far from perfect. In the LEP case, the dominant source of
error is the subtraction of the so called ``$D^{\star\star}$''
contribution to the ${d\Gamma}/{dw}$ distribution.

An alternative approach is to use the measured semileptonic width to charmed
hadrons and relate it to $| V_{cb}|$ through
the Heavy Quark Expansion (HQE) \cite{ikaros}, where the
semileptonic
width is expressed in terms of the $b$ quark mass $m_b$ and a parameter $\mu
_\pi$, related to the average kinetic energy of the $b$ quark moving
inside the $B$ hadron. They obtain:
\begin{equation}
\begin{array}{ll}
| V_{cb}| = & 0.0411 \sqrt{\frac{1.55}{0.105}\Gamma (B\dec X_c \ell \nu)
({\rm ps}^{-1})}\cdot
\left(1-0.025\left(\frac{\mu _{\pi }^2-0.5{\rm GeV}^2}{0.2 {\rm GeV
^2}}\right)\right) \\
~~ & \left[1\pm 0.01|_{m_b}\pm 0.01|_{pert} \pm 0.015|_{1/m_Q^3}\right].\\
\end{array}
\label{vcbinc}
\end{equation}
The semileptonic width depends effectively upon 
$m_b ^2(m_b - m_c)^3$ \cite{ikaros}. Thus 
$|V_{cb}|\approx m_b (m_b-m_c)^{1.5}$.
The $m_b$ errors are taken from the most recent
extractions of the so called `kinetic' $m_b$ (of the order of
1-1.5\%). The term $(m_b-m_c)$ is related to the
spin averaged mass of the $B$ and $D$ mesons:
\begin{equation}
m_b -m_c = <M_B> -<M_D> +\mu _{\pi }^2 \left(
\frac{1}{m_c}-\frac{1}{m_b}\right) +{\cal O}(1/m^2_{b,c}).
\end{equation} 
The uncertainty in $\mu _{\pi }^2$ gives an error of $\pm 0.025$ in
$|V_{cb}|$ and is mostly related to the $m_b-m_c$ dependence upon this
parameter. No error is given for the $1/m^2$
term in eq. 6.
The theoretical errors
quoted above, added linearly, amount to about 6.0\% \cite{ikaros}, a figure 
comparable to the 4.6\% in the 
theoretical extraction with the exclusive method. 
The most recent value of $| V_ {cb}| _{incl}$ combining information
from the four LEP experiments is $(40.76\pm 0.41 \pm 2.0)\times 10^{-3}$. The
first error is the combined statistical and systematic
error in the measurement, added in quadrature, and the latter is the
theoretical error, assumed to be 4.9\%.  An issue that has
gained considerable attention in recent years \cite{nathan} is a possible
sizeable 
source of errors related to the assumption of quark-hadron duality, crucial to
the calculation that lead to eq.~\ref{vcbinc}.
More experimental
checks on the applicability of the quark hadron duality ansatz and
measurements of $m_b$, $m_b-m_c$ and $\mu_{\pi}$ need to be performed 
before we can claim a full understanding of the uncertainties 
in this extraction.

\section{Experimental determination of $| V_{ub} | $}

The hurdles along the path towards a precise determination of the parameter 
$V_ {ub}$ are even more challenging than in the $V_{cb}$ case. The reason is
again deeply rooted in the difficulties that we need to overcome to obtain 
a good estimate of the relevant hadronic matrix elements. In this case
there  is no effective theory like HQET to provide a reliable 
form factor normalization. A variety of
calculations of such form factors exist, based on lattice gauge theory 
\cite{ukqcd}, light cone sum rules  (LCSR) \cite{lcsr},
and quark models \cite{isgw}. Most of them focus on the lightest charmed
hadrons recoiling against the lepton-$\nu$ pair, the $\pi$ and $\rho$ mesons.
In this case they are far from saturating the semileptonic decay width to
charmless hadrons. 
Moreover, because of the light mass of the ``ground state''
hadrons recoiling against the lepton-neutrino pair, a much wider $q^2$ region
is spanned by these decays, adding a strong sensitivity to the $q^2$ 
dependence of the
form factors involved. Lattice gauge calculations are progressing, but they
will produce reliable results for exclusive decays only in
the vicinity of $q^2_{max}$ \cite{sachrajda}.

The first experimental evidence for a non-zero value of the parameter
$|V_{ub}|$ was provided by CLEO \cite{vubcleo}, and soon corroborated by ARGUS
\cite{vubargus}. It was based upon an excess of leptons with momentum
greater than the maximum allowed in the decay $B\dec X_c \ell \nu$. Although 
subsequent CLEO data \cite{endpoint2} provided a very good measurement of
the endpoint
lepton  yield, there are very
convincing theoretical arguments that point to a 
substantial limitation posed by the stringent lepton
momentum cut.  The Operator Product
Expansion (OPE) cannot give reliable predictions because the 
momentum region considered is of the
order of $\Lambda _{QCD}$ and thus an infinite series of terms in this 
expansion may be 
relevant. 

  The CLEO 
collaboration reported the first
convincing evidence for the decays $B\dec \rho \ell \overline{\nu}$ and $B\dec 
\pi \ell \overline{\nu }$ \cite{buprl}. 
CLEO subsequently performed a measurement of the decay $B \dec \rho \ell 
\overline{\nu }$ with a different technique and a bigger
data sample \cite{summerrho}. They used several 
different models to extract 
the value of \vub . Their results are summarized 
in Table \ref{vubrho}.
The first three calculations are based on quark models and their uncertainties 
are guessed to be in the 25-50\%
range in the rate, corresponding to a 12.5-25\% uncertainty for \vub . The other 
approaches, light cone sum rules and lattice
QCD, estimate their errors in the range of 30\%, leading to a 15\% error in
\vub . We can conclude that the average value
of \vub\ extracted with this method is $| V_{ub}| = (3.25 \pm 0.14 ^{+0.21}_{-
0.31}\pm 0.5)\times 10^{-3}$. This corresponds
to a value of $|V_{ub}/V_{cb}| = 0.080 \pm 0.014$. The statistical and 
systematic errors have been added in quadrature
and the theoretical error has been added linearly to be conservative. Note 
that the theoretical error is somewhat arbitrary.

Recently, interest has been stirred by a new approach to 
the extraction of \vub\ 
based on the OPE
approach. The idea is that if the semileptonic width $\Gamma _u$ is 
extracted by 
integrating over the hadronic mass $X_u$ recoiling
against the lepton neutrino pair in a sufficiently large region of phase 
space, the relationship between \vub\ and the 
measured value of the charmless semileptonic branching fraction can be 
reliably predicted. Early estimates (\cite{bdu}, \cite{flw}) showed the
potential of this method and assessed the errors on $|V_{ub}|$ to be of the
order of 10-15\% if a sufficient $X_u$ range was considered. A subsequent
analysis \cite{uraltsev} gave the following assessment of the theoretical 
uncertainties:
\begin{equation}
\begin{array}{ll}
|V_{ub}| =& 0.00442 \left(\frac{Br(B^o\dec X_u \ell \nu}
{0.002}\right)^{0.5}\left(\frac{1.55 {\rm ps}}{\tau _B}\right)^{0.5}\\
~ & \left[1 \pm 0.025_{QCD} \pm 0.035_{m_b}\right] 
\end{array}
\end{equation}
The first error is a lumped estimate of perturbative and non-perturbative 
$QCD$ corrections. Adding the errors linearly, one would assume a 6\% error
in the theory extrapolation. A recent analysis \cite{matthias} favors a more 
conservative but still optimistic 10\% theoretical error.

All the LEP experiments but OPAL attempted 
to use this technique to determine \vub . 
The three experiments combine their analyses and quote 
$|V_{ub}|= (4.13
^{+0.42}_{-0.47}(stat+det)^{+0.43}_{-0.48} (b\dec c) 
^{+0.24}_{-0.25}(b\dec u) \pm 0.02
(\tau _b) \pm 0.20 (HQE)\times 10^{-3}$ \cite{vubwg}.
These measurements have significant $b\dec c$ background
that needs to be understood very well, given the small value of 
$| V_{ub}/V_{cb}|^2$ ($\approx$ 1\%). 
Moreover, predictions to validate the precision of the method 
to measure \vub\ are needed.
The authors of one of the original papers \cite{bdu} include 
an interesting
statement in their abstract: ``$|V_{ub}|$ can be extracted [with the
 method proposed] in a 
largely model-insensitive way. This conclusion is based on the applicability
of the OPE to actual semileptonic $B$ decays. A direct cross-check of this
assumption and a determination of the required basic parameters of the heavy
quark expansion will be possible in the future with more experimental data.'' 
I think that this is a very important program, not yet completed. Important
tests include the extraction of $m_b$ and $\mu_{\pi}^2$ from moments of the
lepton energies and hadron invariant mass in semileptonic decays \cite{fls}. 
Comparing the two sets of values for $m_b$ and  $\mu_{\pi}^2$ from the two
different moment analyses among themselves and with the theoretical
evaluation
would provide an important check. An early preliminary analysis from CLEO
seemed to yield inconsistent results \cite{ichep98}, but we do not have yet
a definitive answer.

Much work needs to be done to achieve a precise measurement of \vub . This
is a quite important element of our strategy to pin down the CKM sector of the 
Standard Model.  
On the theoretical side, large efforts are put in developing more reliable 
methods to determine
the heavy to light form factors. A combination of several 
methods \cite{neub-buras}, all with a limited range of applicability,
seem to be the strategy more likely to succeed. For instance, lattice QCD can 
provide reliable estimates
of the form factors at large momentum transfer, where the discretization errors 
are under control.
HQET predicts a relationship between semileptonic $D$ decays and semileptonic 
$B$ decays. To check 
these predictions and apply them to \vub\ estimates, large data sample with 
reconstructed neutrino momentum 
are necessary.

\section{Conclusions}
This discussion has been focused only on a partial set of the information
used to constrain the $\rho$ and $\eta$ Wolfenstein parameters. My goal is
to urge the community to be cautious in drawing conclusions on the most
probable value of $\rho$ and $\eta$ and their uncertainties obtained
from global fits using averaged quantities with aggressively low errors. 
The present knowledge is a the first important step 
along the way towards  a rich experimental and theoretical program 
involving refinements on these measurements and important additions like
 CP violation observables in
$B$ decays. Much more work is needed to provide a meaningful test
of the Standard Model.

\section{Acknowledgements}
I thank I.I. Bigi, Z. Ligeti and S. Stone for interesting
discussion and K. Ecklund and T. Skwarnicki for useful comments. I 
also want to express my appreciation for the excitement and the
spirit of adventure provided by Y. Rozen to ``Beauty 2000'' and my gratitude
to P. Schlein, who makes each conference in this series a truly remarkable
experience. This work was supported by NSF.

\newpage
\begin{table}
\centering
\begin{tabular}{lll}
\hline 
Experiment & ${\cal F}(1)| V _{cb}| \times 10^3$ 
& $| V _{cb}| \times 10^3 $ \\
\hline
CLEO & $42.4\pm 1.8\pm 1.9$ & $46.4 \pm 2.0 \pm 2.1 \pm 2.1$ \\
\hline
ALEPH & $32.3\pm 2.1 \pm 1.3$ & $35.4 \pm 2.3 \pm 1.4 \pm 1.6$\\
DELPHI & $36.5\pm 1.4 \pm 2.4$ & $40.0 \pm 1.5 \pm 1.8 \pm 1.8 $  \\
OPAL (excl)&$36.6\pm 1.7 \pm 1.8$ & $ 40.1 \pm 1.9 \pm 2.0 \pm 1.8 $\\ 
OPAL (incl)&$37.5\pm 1.3 \pm 2.4$ & $41.1 \pm 1.4 \pm 2.6 \pm 1.9$\\
\hline
LEP AVE &  $34.9\pm 0.7\pm 1.6$ & $38.1 \pm 0.8 \pm 1.8 \pm 1.7 $ \\
\hline
WORLD AVE & $37.0\pm 1.3\pm 0.9$ & $40.5 \pm 1.4 \pm 1.0 \pm 1.8$\\ 
\hline
\end{tabular}
\caption{Summary of $| V _{cb}| $ determinations from the decay
$B\dec D^{\star} \ell \nu$. The parameter $| V _{cb}| $ has been
evaluated using ${\cal F}(1)=0.913 \pm 0.042$ and the last error reflects
the uncertainty in this parameter.}
\label{vcb-excl} 
\end{table}
\newpage
\begin{table}
\centering
\begin{tabular}{ll}
\hline
Model & \vub\ ($\times 10^{-3}$)\\
\hline
UKQCD \cite{ukqcd} & $3.32\pm 0.14 ^{+0.21}_{-0.26}$ \\
LCSR \cite{lcsr} & $3.45 \pm 0.15 ^{+0.22}_{-0.31}$ \\
ISGW2 \cite{isgw} & $ 3.24 \pm 0.14 ^{+0.22}_{-0.29}$ \\
Beyer-Melikhov \cite{bm} & $ 3.32 \pm 0.15 ^{+0.21}_{-0.30}$ \\
Wise/Ligeti \cite{wl} & $2.92 \pm 0.13 ^{+0.19}_{-0.26}$ \\
\hline
Average & $3.25\pm 0.14 ^{+0.21}_{-0.31}\pm 0.5$ \\
\hline
\end{tabular}
\caption{\label{vubrho} Values of \vub\ using  $B\dec \rho \ell \overline{\nu }$ 
and some theoretical models. The \vub\ data
include the results of a recent CLEO analysis \cite{summerrho} and a previous 
CLEO result on exclusive charmless semileptonic
decays \cite{buprl}. The average \vub\ includes an additional contribution 
representative of the theoretical uncertainty
in the measurement.}
\end{table}
\newpage
\end{document}